\documentclass[table]{ecai} 

\usepackage{booktabs}
\usepackage{colortbl}
\usepackage[most]{tcolorbox}
\usepackage{latexsym}
\usepackage{amssymb}
\usepackage{amsmath}
\usepackage{amsthm}
\usepackage{booktabs}
\usepackage{enumitem}
\usepackage{graphicx}
\usepackage{color}

\usepackage{acro}
\usepackage{booktabs}
\usepackage{graphicx}
\usepackage{balance}

\usepackage{titlesec}
\usepackage{hyperref}
\usepackage{multirow}
\usepackage{siunitx}
\usepackage{subcaption}

\usepackage{graphicx}
\usepackage{svg}
\usepackage{booktabs}
\usepackage{textcomp}
\usepackage{algorithm}
\usepackage[noend]{algpseudocode}
\usepackage{float} 
\usepackage{array}
\usepackage{soul}
\usepackage{tikz}
\usepackage{comment}

\usepackage{graphicx}

\newcommand{\BibTeX}{B\kern-.05em{\sc i\kern-.025em b}\kern-.08em\TeX}

\begin{document}


\begin{frontmatter}


\paperid{123} 


\title{To Judge or not to Judge: Using LLM Judgements for Advertiser Keyphrase Relevance at eBay}


\author
{\fnms{Soumik}~\snm{Dey}\orcid{0009-0006-9763-318X}\thanks{Corresponding Author. Email: sodey@ebay.com}}
\author
{\fnms{Hansi}~\snm{Wu}\orcid{0009-0004-5520-9079}}
\author
{\fnms{Binbin}~\snm{Li}\orcid{0009-0005-3131-3995}} 

\address{eBay Advertising \\
eBay Inc. \\
San Jose, California, USA}


\begin{abstract}
E-commerce sellers are recommended keyphrases based on their inventory on which they advertise to increase buyer engagement (clicks/sales). The relevance of advertiser keyphrases plays an important role in preventing the inundation of search systems with numerous irrelevant items that compete for attention in auctions, in addition to maintaining a healthy seller perception. In this work, we describe the shortcomings of training Advertiser keyphrase relevance filter models on click/sales/search relevance signals and the importance of aligning with human judgment, as sellers have the power to adopt or reject said keyphrase recommendations. In this study, we frame Advertiser keyphrase relevance as a complex interaction between 3 dynamical systems --- seller judgment, which influences seller adoption of our product, Advertising, which provides the keyphrases to bid on, and Search, who holds the auctions for the same keyphrases. This study discusses the practicalities of using human judgment via a case study at eBay Advertising and demonstrates that using LLM-as-a-judge en-masse as a scalable proxy for seller judgment to train our relevance models achieves a better harmony across the three systems, provided that they are bound by a meticulous evaluation framework grounded in business metrics.

\end{abstract}

\end{frontmatter}

\section{Introduction}

\label{sec:Introduction}

\begin{figure*}[t]
\centering
\begin{subfigure}{8.5cm}
\centering
\includegraphics[width=\linewidth]{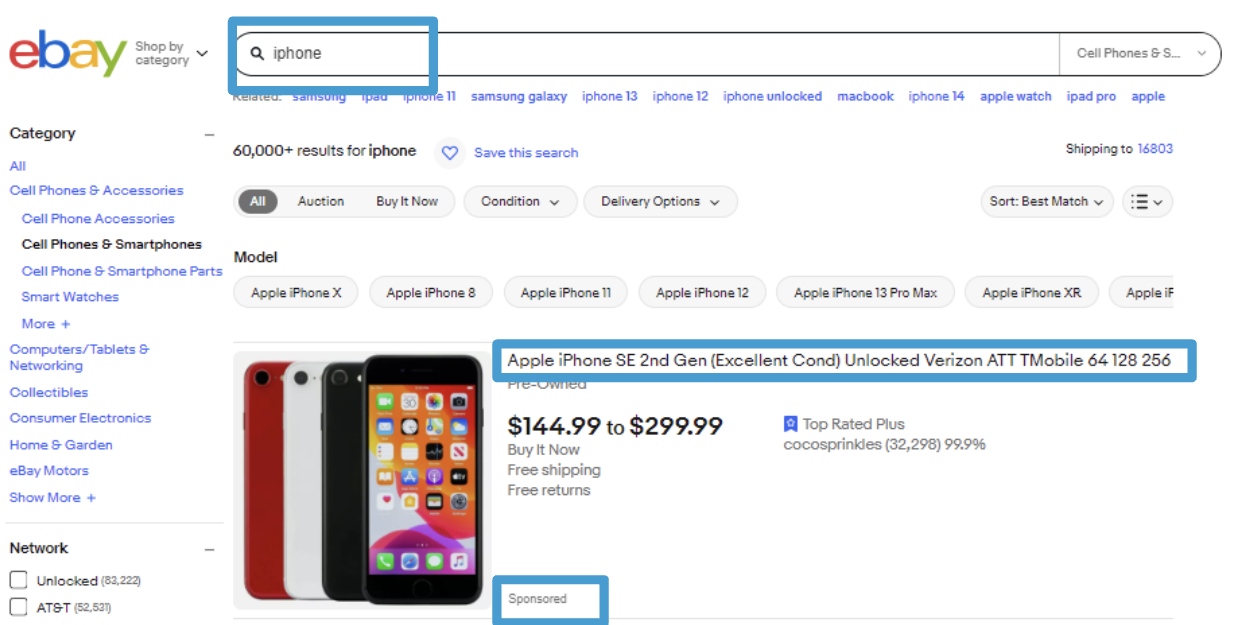}
  \caption{Buyer side}
  \label{fig:1a}
\end{subfigure}\qquad
\begin{subfigure}{8.5cm}
\centering
\includegraphics[width=\linewidth]{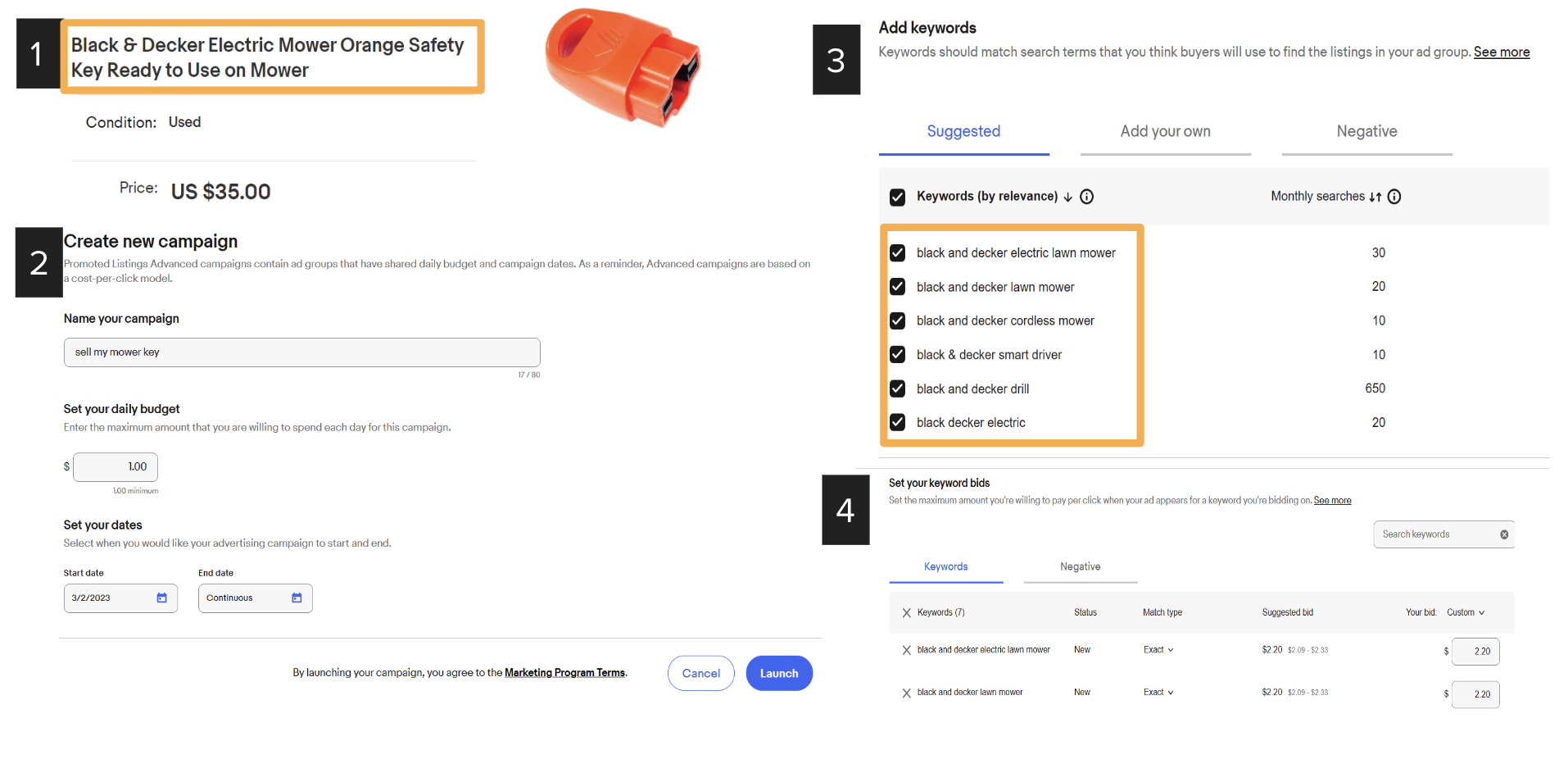}
\caption{Seller Side}
\label{fig:1b}
\end{subfigure}
\vspace{4mm}
\caption{Screenshot of our keyphrases for manual targeting in Promoted Listings Priority~\cite{eBay} for eBay Advertising.}
\vspace{4mm}
\label{fig:screenshot}
\end{figure*}

\begin{figure*}[t]
    \includegraphics[width=\linewidth]{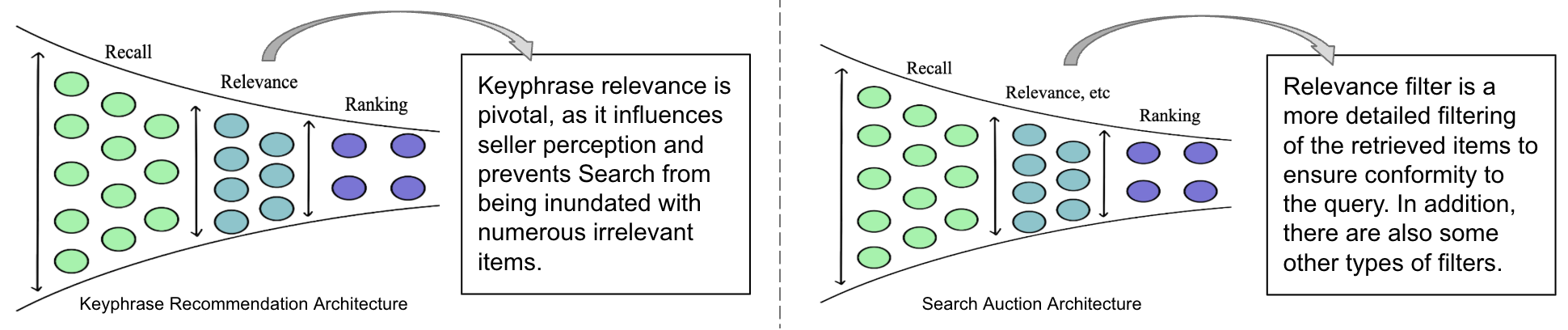}
\caption{A side by side comparison of eBay Advertising for Keyphrase Recommendation funnel and eBay Search funnel for Auctions.}
\vspace{2mm}
\label{fig:funnel}
\end{figure*}

In the realm of e-commerce, sellers adeptly employ online advertising solutions, including the use of keyphrase recommendations, to counteract their often suboptimal positions in organic search results. This approach enables them to secure a beneficial presence on the search results page (SRP) and cultivate a greater engagement with potential buyers, see Figure \ref{fig:screenshot}. At eBay, the process of Advertiser keyphrase recommendation comprises three fundamental steps: \textit{retrieval}, which entails the aggregation of keyphrases that are pertinent to the product; \textit{relevance}, which involves the further refinement of these keyphrases to ascertain their pertinence to the item in question; and \textit{ranking}, which prioritizes the adoption of keyphrases based on the seller's budgetary constraints and preferences. The role of keyphrase relevance is of paramount importance, as it shapes seller perceptions and aids in preventing the inundation of search systems with numerous irrelevant items vying for attention in auctions. Advertiser keyphrase relevance models are trained to discern overarching patterns within click and sales data. A keyphrase that has attracted a significant amount of clicks or sales relative to an item can be deemed relevant to that item. Fundamentally, clicks and sales constitute robust affirmative indicators of relevance; however, they are not effective as indicators of non-relevance. E-commerce datasets are susceptible to missing-not-at-random (MNAR) conditions due to various biases ~\cite{surveybias, sampleselectionbias, clicksNquery, beyondPosBias, learning2rank, debiasedness, MNAR, lim2015top}. The absence of clicks for an item in relation to a specific query does not inherently imply the item's irrelevance. Within the e-commerce framework, buyers function as annotators, yet, unlike traditional annotators, they are exposed to a biased presentation of items as a result of search rankings, which in turn influences their annotations in terms of clicks or sales. Consequently, an item that lacks popularity will occupy a suboptimal position on the SRP because search rankings are derived from clicks, and as a result, it may fail to secure any clicks or sales. This bias undermines the reliability of negative relevance signals that are based solely on clicks or sales.

\begin{figure*}[t]
\centering
\includegraphics[width=\linewidth]{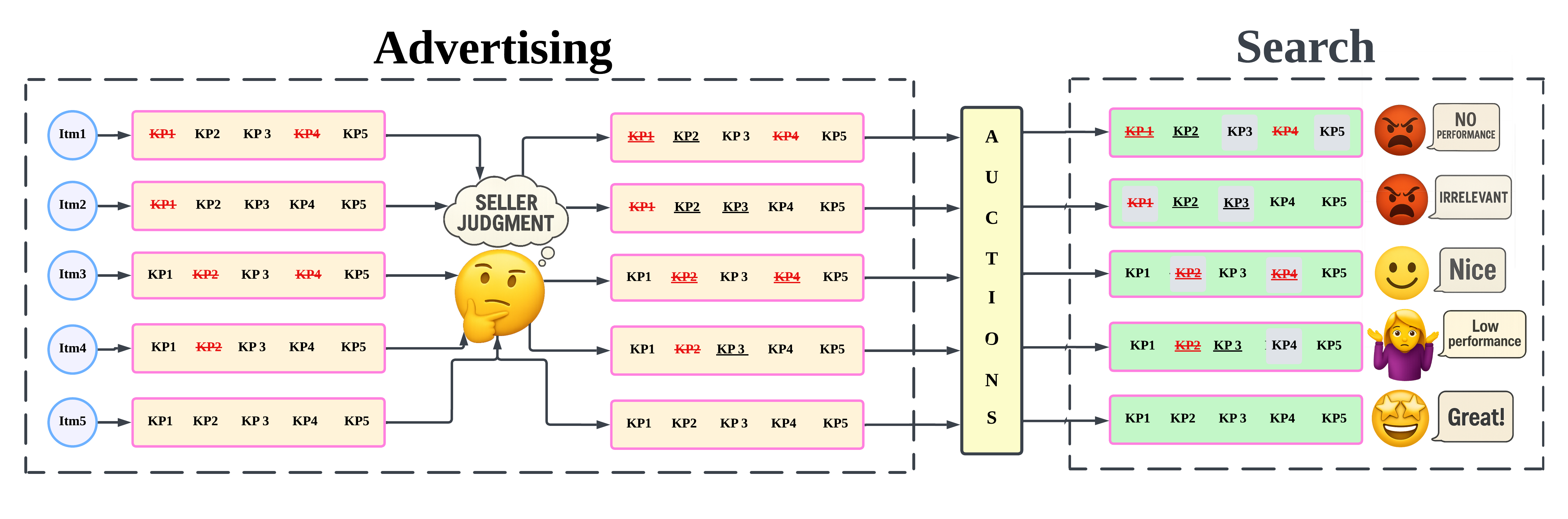}
\caption{Auction mechanism of items (Itm) in relation to keyphrases (KP). Red strikethrough font represents filter of Advertising, the underline represents seller curation of keyphrases after advertising has filtered them while gray highlight represents the relevance filter of Search. }
\label{fig:auctions}
\vspace{2mm}
\end{figure*}  

 \textit{eBay Search} shows items in response to buyer queries and also involves the same tasks, see Figure \ref{fig:funnel}: \textit{retrieval}, retrieving items related to the search queries; \textit{relevance}, a more detailed filtering of the retrieved items to ensure conformity to the query, and \textit{ranking}, i.e. ordering the retrieved items based on potential to achieve clicks so that buyers interact with them preferentially. From the sellers' perspective, when \textit{eBay Advertising} presents advertised keyphrases to sellers, they choose to bid on them for their items. These items then enter an auction corresponding to the keyphrase. This auction process involves a complex interaction with \textit{eBay Search}. During this auction, \textit{eBay Search} acts as a \textit{middleman} and aligns the items with the queries, which are exactly the advertised keyphrases recommended by eBay Advertising, and further filters the auction items using Search's relevance filter. This auction mechanism also ensures that the click data logged will only contain keyphrases which pass Search's relevance filter (auction winning impression gaining keyphrases are logged). \textit{This introduces an additional bias --- training will only be on keyphrases which pass the Search relevance filter}. Training on click data ensures that the model never experiences keyphrases that are irrelevant to Search, while Advertising does generate such keyphrases and needs to filter them. This \textit{middleman bias} is a form of sample selection bias~\cite{rec4ad, sample_selection}, which the click data additionally suffers from in the context of advertiser keyphrase recommendation.

The interaction between the Sellers, Advertising and Search displays an inherent imbalance --- Sellers have the ultimate say in adopting Advertising's keyphrases while Search has the ultimate say in rejecting those keyphrases at the time of auction. For instance, take \texttt{Itm1} from Figure \ref{fig:auctions}; Advertising considers \texttt{KP1} and \texttt{KP4} irrelevant, while \textit{eBay Search} regards \texttt{KP2}, \texttt{KP3}, and \texttt{KP5} as irrelevant and the seller rejecting KP2 as a legit keyphrase, leading to \texttt{Itm1} not participating in any auctions. \textit{This outcome occurs regardless of the true relevance of the keyphrases, which means that even if \texttt{KP2}, \texttt{KP3}, and \texttt{KP5} are genuinely relevant, \texttt{Itm1} will not enter those auctions because the seller and eBay Search do not find them relevant}. An ideal scenario of complete agreement would be for \texttt{Itm5} and \texttt{Itm3} where all parties are in complete agreement. \footnote{For \texttt{Itm3} the outcome is not as great as \texttt{Itm5} as there are still some keyphrases which are not getting into auctions. However, that is more of a retrieval problem than a relevance problem, i.e., why are the keyphrases \texttt{KP2}and \texttt{KP4} getting recommended in the first place?}

Understanding the dynamics and mechanisms of auctions is essential for the successful execution of advertising campaigns. However, this understanding is further complicated by the complex nature of sellers’ perceptions. Promoted Listings Priority, our product, enables manual targeting, which allows sellers to select their preferred keyphrases while also granting them the option to follow our recommendations. For instance, should sellers judge \texttt{KP2} for \texttt{Itm1} as irrelevant, they may choose to disregard this particular keyphrase, thereby rendering our suggestions ineffective. This scenario can escalate, as sellers often encounter an abundance of "irrelevant" keyphrases, discouraging them from proceeding with the adoption of our product. Such outcomes not only lead to diminished Seller Satisfaction scores, but also represent a wasted utilization of resources and opportunities for eBay. Moreover, offering ostensibly \textit{"sensible"} keyphrases lacking substantial targeting capabilities can result in subpar advertiser campaign performance, potentially prompting sellers to leave the platform. Consequently, it is imperative to ensure that keyphrases resonate with human judgment while simultaneously maintaining robust performance. In this study, we interpret Advertiser keyphrase relevance as a multifaceted interplay among three dynamical systems: seller judgment impacting their product adoption, Advertising supplying the bid and the keyphrases, and Search orchestrating auctions for those keyphrases.

\section{Previous Experiments and Related Work}

This section highlights prior experiments, their limitations, and our insights. An earlier BERT model, based on human judgment and 80,000 records with 3 annotators, encountered issues post-launch; sellers reported irrelevant keyphrases. A post-launch analysis showed insufficient category representation due to eBay's extensive categories, necessitating far more data than 80,000 records.\footnote{The vast scope of eBay's 100,000+ categories demands substantial data for adequate coverage.} An xgboost Click-Through rate (CTR) model using click data was also tested but failed offline evaluations for search alignment and spot-checked human assessments. A Jaccard \cite{jaccard} index-based model, a rule-driven token algorithm with a relevance filter, outperformed and was deployed. The CTR model included the Jaccard index as a feature but still fell short. Further investigation revealed the CTR model's significant bias, lacking representation of item-query pairs that failed search relevance, perpetuating e-Commerce data biases discussed in Section \ref{sec:Introduction}.


The Jaccard index, computed as the ratio of token intersection to union, encountered specific challenges. It disproportionately penalized shorter keyphrases, which are generally head keyphrases critical in driving major engagement—10\% of keyphrases while accounting for 90\% of clicks and sales on eBay. The introduction of the Jaccard filter resulted in the exclusion of single-token keyphrases, key elements in advertising for sellers. Furthermore, the lack of semantic awareness in Jaccard meant it often missed keyphrases that were semantically relevant to listings. While straightforward to implement within eBay systems, its utility as a relevance model is limited and primarily served as a temporary solution.

In the study documented in \cite{dey2025middlemanbiasadvertisingaligning}, we then devised a 2-layer cross-encoder model that was trained utilizing search relevance scores. This was specifically aimed at overcoming the intermediary biases present in click-based training data, while simultaneously establishing a more resilient framework compared to the traditional rule-based Jaccard model. This cross-encoder, by capitalizing on search relevance indicators, demonstrated superior performance relative to the Jaccard relevance model, largely by reducing the inherent rule-based biases that disadvantage shorter keyphrases (potentially head keyphrases), within the Jaccard model. Despite these improvements, it is critical to note that this model did not entirely align with human evaluations. The perception held by sellers is paramount for the successful adoption of our product, as dissatisfaction among sellers significantly impacted the perceived relevancy of our keyphrases, subsequently reflected in the Seller Sentiment scores. \cite{dey2025middlemanbiasadvertisingaligning} neglected to adequately consider the essential role of seller perception and its influence on our keyphrase recommendation strategies. Hence, there was a pronounced need for an innovative model that more accurately aligns with human judgment while ensuring precise targeting capabilities without resulting in diminishing returns for the sellers.

Within the realm of ongoing research, as detailed in \cite{wang2024improvingpinterestsearchrelevance, gurjar2025dashclipleveragingmultimodalmodels, microsoft_llm_as_a_judge, gu2025surveyllmasajudge}, researchers have investigated the innovative application of augmenting search data through the integration of labels generated by Large Language Models (LLMs) for the development of advanced relevance and retrieval models. Their analysis particularly emphasizes the scalable application of LLM-generated labels, which effectively addresses the significant expenses associated with collecting data labeled by human annotators. The researchers initiated their approach with a limited dataset comprising human annotations, which was utilized to fine-tune an LLM. This fine-tuned LLM subsequently facilitated the production of a more extensive array of labels, which in turn were employed in the training of a more compact model, ultimately operationalized in a production environment. Despite the growing inclination towards employing LLMs not only as evaluative tools but also as data augmentation tools, the literature concurrently presents a body of criticism \cite{bavaresco2024llmsinsteadhumanjudges, Soboroff_2025, stureborg2024largelanguagemodelsinconsistent}. These critiques examine the appropriateness of using LLMs as evaluators and data generators and explore the multifaceted considerations pertinent to their roles in such capacities.

There are generally two main methods to employ LLM-as-a-judge for data generation/augmentation:

\begin{itemize}
    \item \textbf{General LLM:} For successful LLM-as-a-Judge automation, advanced models like GPT-4 \cite{NEURIPS2022_b1efde53, openai2024gpt4technicalreport} provide a feasible substitute for human assessments \cite{NEURIPS2023_91f18a12}. Li et al. \cite{alpaca_eval} developed an 805-question test set to evaluate performance against text-davinci-003 using GPT-4. Zheng et al. \cite{zhu2025judgelmfinetunedlargelanguage} created 80 multi-round questions across eight sectors, employing GPT-4 for automated scoring. Despite GPT-4's demonstrated accuracy and consistency over human experts, usage is often hindered by rate limits or API call restrictions to external sources. \cite{ashirbad-etal-2024, zhang2024lazyprolifictacklingmissing} utilized the Mixtral-7x8B Instruct v0.1 \cite{jiang2024mixtralexperts} model to annotate keyphrase relevance in advertising effectively. Its open-source status also aids in model distillation and training. \footnote{During its development, models such as LLAMA 2 \cite{touvron2023llama2openfoundation}, DBRX \cite{databricksIntroducingDBRX}, and Qwen-2 \cite{yang2024qwen2technicalreport} were considered but faced distillation and licensing constraints for commercial use.} A general LLM's weakness in instruction adherence or reasoning could undermine the LLM-as-a-Judge methodology, whereas its generality of being a World Knowledge recommender \cite{lichtenberg2024largelanguagemodelsrecommender} makes it immune to other biases that fine-tuned models might be subject to.
    \item \textbf{Fine-tuned LLM:} Fine-tuning a judge model typically include \cite{huang2024empiricalstudyllmasajudgellm}: (1) Data Collection—comprising instructions, evaluation subjects, and assessments, often using instruction datasets and GPT-4 or human annotations. (2) Prompt Design—adapting prompt templates to the evaluation method. (3) Model Fine-Tuning—applying designed prompts and data to fine-tune the model using the instruction tuning framework. The model uses instructions and responses to generate evaluation outputs. Post-fine-tuning, the evaluator model can assess target objects. Despite improved test set performance, these models face limitations in evaluation abilities and retained biases from human annotators \cite{gu2025surveyllmasajudge}. Other issues like improper design of prompts and datasets can lead to poor generalization, complicating comparisons with strong models like GPT-4.
\end{itemize}
 
Moreover, the assessment of models developed using data where LLMs serve as judges reveals a notable trend: these models are frequently evaluated by the LLMs themselves. This practice precipitates a self-enhancement bias—a situation wherein the LLM evaluators exhibit a preference for responses they have generated. Given the considerable concerns surrounding self-enhancement bias, as highlighted in \cite{ye2024justiceprejudicequantifyingbiases}, it is advisable to refrain from employing the identical model for evaluation purposes. Nonetheless, this presents only a temporary measure, as it may lead to the selection of non-optimal evaluators when appraising the most sophisticated LLMs. Evaluators grounded in some sort of business impact measures would be a better alternative.

\section{Experiment Design}

\subsection{Dataset Collection}

\subsubsection{Human Judgment}

\begin{figure}[t]
\includegraphics[width = \linewidth]{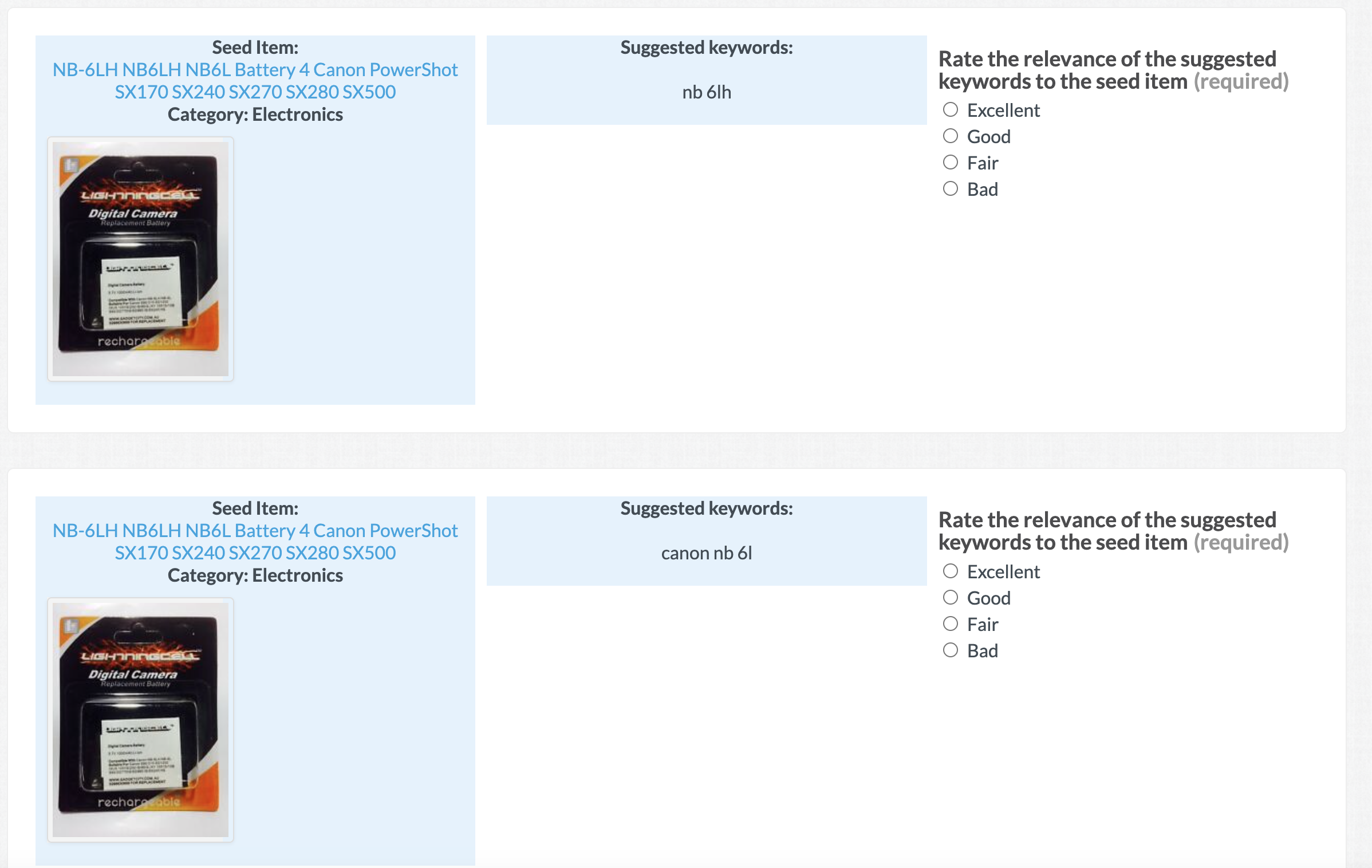}
\vspace{0.1mm}
\caption{Interface for our human annotators.}
\vspace{5mm}
\label{human}
\end{figure}

In conducting human relevance evaluations, we acquire crowd-sourced data concerning the assessment of our keyphrase suggestions. Annotators are provided with the title of the item, an associated image, and the recommended keyphrase. They are tasked with evaluating the relevance of our keyphrases in relation to the item, assigning them labels from a predefined scale (excellent, good, fair, bad). The instructions given to the annotators are given below while the interface is illustrated in Figure \ref{human}.

\begin{tcolorbox}[colback=gray!5!white, colframe=black, title=Ads Keyword Suggestion for Evaluation]

\footnotesize{
Sellers participating in the Promoted Listings ads program can bid on search query keywords and phrases where they want their items to appear. Help us to evaluate the performance of an automated tool that suggests keyword phrases for which a seller might want their ad to appear. How relevant are the suggested keywords to the item?

\vspace{0.5em}

\textbf{Question: How relevant are the suggested keywords to the seed item?}

\begin{itemize}[leftmargin=*]
    \item \textbf{Excellent} -- the suggested keywords are highly relevant and match the seed item in all core attributes.
    \item \textbf{Good} -- the suggested keywords are somewhat relevant and provide an OK or good enough match to the item. Some of the secondary attributes might not match, but important traits like model, size, or product type are respected.
    \begin{itemize}
        \item Note, for the purposes of ad placement, keywords containing a competing brand and similar model can qualify as Good.
        \item Example: seed item is a Samsung Galaxy phone, suggested KWs are ``google pixel pro.'' Both \textbf{phones} run on Android and are flagship models so have enough similarity to qualify as Good.
    \end{itemize}
    \item \textbf{Fair} -- the suggested keywords are only slightly relevant and miss the target in an important way. Core traits like product type or size might not match, but you can understand the connection between the seed item and suggested keywords.
    \item \textbf{Bad} -- the suggested keywords are not at all relevant and do not match the seed item on most, if not all, significant traits.

\end{itemize}
}
\end{tcolorbox}

We amassed approximately 150,000 annotations regarding our keyphrase suggestions. Each pairing of item and keyphrase is independently evaluated by three annotators, after which a conclusive judgment is reached. Only assessments that demonstrate total agreement among the annotators are retained for the purposes of this study. Given the binary nature of our classification task (i.e., relevant versus irrelevant), annotations labeled excellent or good are categorized as relevant, whereas those marked as bad or fair are reclassified as irrelevant. \footnote{Given the constraints of budget, time, and an inherited framework of graded human assessment not originally designed for this task, we resorted to converting to binary relevance in the absence of more suitable options.} 

\subsubsection{Click Data}

Since clicks are a strong indicator for relevance, we gather click data amounting to roughly 24 million item keyphrase pairs over a period of 30 days with at least 30 impressions and 1 clicks with a minimum CTR of 0.1. Click data can be seen as data annotated by millions of buyers, and it is a stronger signal as buyers are spending their time to do so. The minimum thresholds of CTR, impressions and clicks are used to eliminate noise such as 1 click 1 impression or 1 click and 1000 impression. An important thing to note is that due to the nature of the auction process, all click data is deemed relevant by eBay Search (i.e., they are only surfaced to buyers because it is deemed relevant by eBay Search).

Seller adoption of our keyphrases is logged as well --- however, they are unreliable indicators of negative judgment as to the way they are presented, i.e., they are presented along with the bid. It is hard to disambiguate the exact cause of non-adoption --- is it bids, or keyphrases? While a case can be made for positive seller adoption as a reference for positive human judgment, it is redundant in the sense that adopted keyphrases accepted by sellers will anyway be reflected in click data if they receive clicks. Hence, we use click data because it is a stronger signal that is aligned with both the buyer and the seller judgment.  

\subsubsection{Search Relevance Data}

As detailed in \cite{dey2025middlemanbiasadvertisingaligning} the usage of search relevance data is a good way to mitigate \textit{middleman bias} in Advertising. The Search relevance data can act as a strong indicator of irrelevance due to nature of auctions illustrated in Figure \ref{fig:auctions} --- Search has the ultimate say in who gets to participate in auctions. We collected around 50 million search relevance scores over a period of 5 months on our keyphrase recommendations.

\subsubsection{LLM Judgment Data}

In order to obtain a set of scores from a large language model (LLM) as either an augmentation or a substitute for human evaluation, we collected 50 million judgment scores derived from the Mixtral 8x7B Instruct v0.1 model. This process utilized two distinct versions of the Mixtral 8x7B Instruct v0.1: 1) the standard general-purpose LLM, and 2) a version fine-tuned specifically to binary human judgment labels. The general-purpose LLM demonstrates a 90\% concordance with click data, which serves as an indicator of positive human judgments. However, when evaluated against the human judgment scores collected in the course of our analysis, the standard LLM shows a fair level of agreement, with a kappa coefficient of 0.258. In contrast, the fine-tuned LLM exhibits superior alignment with human-provided judgments, achieving a kappa coefficient of 0.724, and shows 95\% concordance with click data. The prompt applied in this study is provided below.

\begin{tcolorbox}[colback=gray!5!white, colframe=black, title=Prompt Design]
\noindent\texttt{\footnotesize{Below is an instruction that describes a task. Write a response that appropriately completes the request. \\
\\
\#\#\# Instruction: \\
Given an item with title: "\{\textrm{title}\}", determine whether the keyphrase: "\{\textrm{keyphrase}\}", is relevant for cpc targeting or not by giving ONLY yes or no answer: \\
\\
\#\#\# Response:
}}
\end{tcolorbox}

\subsection{Modeling}

We used the data sources described in the previous section to model a cross encoder on cross-entropy loss with \texttt{keyword [SEP] category name [SEP] item title} as the format of input. Not all data sources are created equal --- due to the limited nature of the human judgment data it is merely used as a dataset to create the fine-tuned LLM model whereas the click data is only used for a positive signal and has to be combined with other sources of data to train a binary classifier (relevant/irrelevant).

We trained 5 models to compare with the production model. The models are described below: 

\begin{itemize}
    \item \textbf{Search (+/-)}: This particular model underwent training using the search relevance data, which is actively utilized in the current production environment and thoroughly explained in \cite{dey2025middlemanbiasadvertisingaligning}. Within this context, the search relevance labels serve as the definitive truth markers, delineating both relevant and irrelevant data, for the purpose of training the binary classification algorithm.
    \item \textbf{Click (+) Search (-)}: This model received training on datasets comprising positive click data and negative search data. Essentially, it can be regarded as a model exclusively trained on indicators pertinent to business outcomes, with the click data representing clear-cut positive confirmations and the negative search relevance data serving as definitive negative indicators.
    \item \textbf{Click (+) LLM (+/-)}: In this model's training, click data provided unequivocal positive evidence, while the LLM data served as a surrogate to represent both affirmative and negative human evaluations.
    \item \textbf{Click (+) LLM (-)}: This model was formulated by incorporating click data serving as incontrovertible positive evidence, supplemented by negative data from the LLM. The positive data from the LLM was discarded, with the remaining data acting as a stand-in for adverse human evaluation.
    \item \textbf{LLM (+/-)}: The exclusive training of this model utilized relevance labels produced by the general Mixtral 8x7B Instruct v0.1 model, where assessments from the LLM provided signals both positive and negative relevance.
    \item \textbf{fine-tuned LLM (+/-)}: The training of this model relied solely on relevance labels generated by the fine-tuned version of the Mixtral 8x7B Instruct v0.1 model, fine-tuned using human evaluation data, thus enabling the LLM assessments to serve as proxies for both positive and negative human judgment.
\end{itemize}

Our chosen cross-encoder is a compact 2-layer \texttt{bert-tiny} \cite{bert-tiny, bhargava2021generalization} model trained using Cross-Entropy loss on the binary labels, selected due to its efficient execution and demonstrated effectiveness within our system. \footnote{Refer to the earlier study in \cite{dey2025middlemanbiasadvertisingaligning}, which illustrates that 2-layer cross-encoders outperform larger bi-encoders with only a marginal increase in latency.} The training process of the cross-encoder utilized 4 Nvidia A100-80GB GPUs, over the course of 8 epochs, implementing a learning rate of 2e-4, and deploying a per-device batch size of 40960. \footnote{The cross-encoder was additionally trained using MSE Loss on the yes/no token probabilities derived from the LLMs, but the performance was significantly poorer, therefore we omit the results.}

\subsection{Evaluation framework}

Given the diverse origins and varying volumes of our data sources, we opted to test the effect of multiple data sets. Due to the heterogeneous characteristics of these datasets and the inherent self-enhancement bias observed in large language models (LLMs), relying solely on LLMs for result evaluation is problematic. Therefore, it is essential to benchmark evaluations using certain business metrics, which can be simulated, or are attainable in an offline environment. Subsequently, the most suitable candidate model can be tailored to meet specific business needs and subjected to an online A/B testing process.

In order to assess the performance of the general large language model (LLM) framework as described in \cite{mishra2024graphexgraphbasedextractionmethod}, we used click data as a reliable benchmark. This data serves as a strong indicator of positive relevance, contributing to a dependable evaluation process. The LLM utilized in this study Mixtral 8x7B Instruct v0.1, exhibited substantial alignment with a comprehensive array of human-generated positive annotations, achieving an impressive over 90\% agreement by providing affirmative responses in the evaluated click data. This achievement underscores its robust performance in realistic settings. In our own research, our aim is to adopt a methodology that reflects this benchmarking framework. To scrutinize the effectiveness of the proposed candidate models, our initial step involved calculating threshold values for each model that ensure the retention of 95\% of clicks extracted from search logs. \footnote{The stipulated 95\% threshold is derived from business logic criteria and is currently implemented for the existing production models.} Utilizing these threshold determinations, we then evaluated three primary aspects: 1) the potential loss of sales revenue when applying these thresholds to identical click data, 2) the reduction in keyphrases when the same thresholds are applied to a segment of our keyphrase recommendations and, 3) the \textit{search pass rate}, i.e. the percentage of the keyphrases that after passing the Advertising relevance filter pass the search relevance filter. An optimal model is characterized by minimizing sales loss while maintaining consistent click performance and simultaneously reducing a larger fraction of our recommended keyphrases and maintaining a healthy search pass rate for efficiency.

\definecolor{green1}{RGB}{30,70,30}     
\definecolor{green2}{RGB}{80,130,80}
\definecolor{green3}{RGB}{140,190,140}
\definecolor{green4}{RGB}{200,230,200}
\definecolor{green5}{RGB}{235,250,235}  

\begin{table}[t]
\centering
\begin{tabular}{c||ccc}
\rowcolor{gray!30}
\toprule
\textbf{Model} & \textbf{\# Keyphrases} & \textbf{Sales} & \textbf{Search Pass Rate}\\
\midrule
Search (+/-)               & \cellcolor{green3}{0\%}     & \cellcolor{green2}{\textcolor{white}{0\%}} & \cellcolor{green5}{0\%} \\
Click (+) LLM (-)  & \cellcolor{green2}{\textcolor{white}{-31\%}}  & \cellcolor{green4}{-48\%} & \cellcolor{green1}{\textcolor{white}{+103\%}} \\
LLM (+/-)             & \cellcolor{green1}{\textcolor{white}{-68\%}}  & \cellcolor{green1}{\textcolor{white}{+10\%}} & \cellcolor{green3}{+97\%} \\
Click (+) LLM (+/-)      & \cellcolor{green2}{\textcolor{white}{-55\%}}  & \cellcolor{green3}{-23\%} & \cellcolor{green2}{\textcolor{white}{+100\%}} \\
Click (+) Search (-)    & \cellcolor{green3}{-26\%}                     & \cellcolor{green5}{-60\%} & \cellcolor{green3}{+82\%} \\
fine-tuned LLM (+/-)             & \cellcolor{green5}{+7\%}   & \cellcolor{green3}{-20\%} & \cellcolor{green4}{+56\%} \\
\bottomrule
\end{tabular}
\vspace{1mm}
\caption{Table showing the performance of all models in comparison to the production model Search (+/-). The numbers show the increase/decrease in the volume of our recommendations normalized by item count and the sales loss/gain in comparison to the production model when the threshold for maintaining 95\% clicks is applied and the percentage of keyphrases that pass the search relevance filter. Darker hues denote more advantageous outcomes.}
\label{tab:Results}
\vspace{4mm}
\end{table}

\section{Results}
\label{sec:results}

\subsection{Offline Results}

Analyzing the data presented in Table \ref{tab:Results}, it is evident that the most optimal outcomes are achieved with the LLM (+/-) model. The Click (+) LLM (+/-) emerges as a close contender in terms of performance. Notably, the LLM (+/-) model surpasses the production model by eliminating a significantly higher number of recommendations, a decrease by 68\%, while simultaneously achieving an increase in sales by 10\%. Interestingly, the fine-tuned LLM (+/-) model demonstrated the least desirable results, exhibiting a particularly poor performance relative to other models under consideration. Paradoxically, it increased the frequency of keyphrase recommendations while simultaneously negatively impacting sales. This finding contrasts with existing research where LLMs, initially fine-tuned based on human evaluations, tend to be more successful either as evaluators or as generators of initial data. In addition, all models, with the exception of LLM (+/-), exhibited a decline in sales, indicative of inefficiency in both their filtering and learning processes. All models see significant improvement in their search pass rate, in comparison to the production model. The production model suffers from the thresholding to maintain clicks, resulting in lowering of thresholds and giving away too many search irrelevant keyphrases. From the search pass rate perspective, the fine-tuned LLM again fares the second worst, with the Click (+) LLM (-) and Click (+) LLM (+/-) models being the best and the second best, respectively. The LLM (+/-) model does do relatively well showing that Search and LLM judgments are well aligned.

\subsection{Error Analysis}

Oftentimes, there is a tendency to adopt or reuse human judgment frameworks across different projects or use cases in the industry. Our experiments provide a cautionary tale in doing so. Since our results ran counterintuitive to the norm of fine-tuned LLMs serving as data augmentors or evaluators, we did post-experimental analysis as to the reason behind this observation. We could identify potential problems with our human judgement data:

\begin{itemize}
    \item Annotators were asked to rate pairs of keyphrases of items with four distinct labels --- \textit{excellent}, \textit{good}, \textit{fair} and \textit{bad}. While they were given instructions as to \textit{excellent} or \textit{good} being extremely relevant and somewhat relevant respectively while \textit{fair} and \textit{bad} being bad targeting and irrelevant respectively --- the open-ended interpretation and the complex nature of such labels might have been detrimental to the annotators' judgment. Instead, a simple binary assessment would have been more appropriate as our final goal is to do a binary classification.
    \item The scale of annotations might have been inadequate given the huge inventory of items on eBay and the diverse set of buyer queries and activity we see everyday.
    \item In addition, while annotators were rendering the judgment, they were provided a picture of the item in question while the models are not multimodal and are not given any image as input. This creates a problem of modality bias \cite{park2024assessingmodalitybiasvideo, 10.1145/3565266, dey24_speechprosody} --- both the annotators and the model should have access to the same modalities (sources of input) while they are making a judgment. An example is for an item like \texttt{``iPhone 11 64GB 128G Unlocked ATT Boost Cricket Spectrum Excellent Condition''} which is yellow in color presented in the image but not in the title in any way and keyphrases \texttt{``yellow iphone''} and \texttt{``red iphone''} the annotator would rate the first as relevant and the second as irrelevant. But when the downstream cross-encoder model is presented with this information for distillation, with the absence of visual clues to guide the model, these data points serve as noise misguiding the model. 
\end{itemize}

Although it is crucial to develop enhanced strategies for acquiring human judgments, devising unique designs for every individual use case proves to be impractical. Considering the vastness of web data, coupled with our platform managing 2.3 billion items across a diverse array of products and use cases, alongside the expense associated with obtaining human evaluations, it becomes impractical to tailor designs for each specific case. Additionally, when large language models (LLMs) are fine-tuned on such data, the biases intrinsic to the dataset tend to be magnified and propagated to subsequent models, thus leading to erroneous outcomes. In these scenarios, a meticulous assessment is essential to evaluate the data's usage and the necessity of employing a fine-tuned LLM. Indeed, for practical purposes, it may be advantageous to utilize a general-purpose LLM, provided that a comprehensive evaluation framework, based on business metrics, is meticulously implemented.

Moreover, the distillation procedure and click thresholding present challenges across various models. For instance, the item \texttt{``Genuine 15V 4A Power AC Adapter Laptop Charger For Surface Pro 3 4 5 6''} and the keyphrase \texttt{``microsoft surface charger''} are considered relevant by both LLMs; however, the distilled model fine-tuned LLM (+/-) does not, despite the original LLMs' agreement. The cross-encoder distilled for the fine-tuned model displays lower calibration accuracy, excluding high-conversion items to meet a 95\% click threshold. Conversely, distilled models from alternate sources show superior calibration, effectively preserving clicks and sales.

\begin{figure*}[t]
\centering
\includegraphics[width=0.8\linewidth]{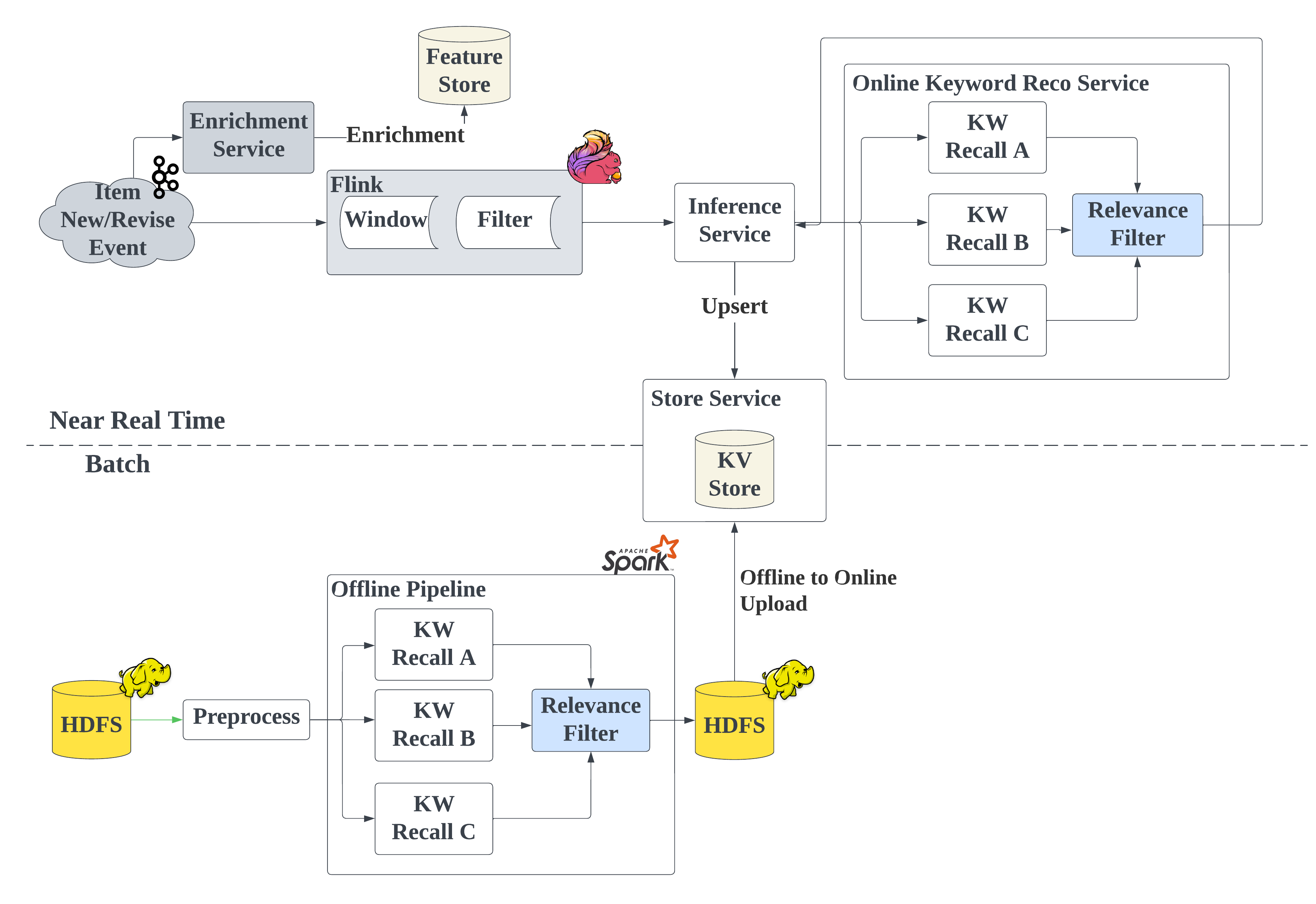}
\caption{Production Serving architecture for our retrieval and relevance models.}
\label{fig:prod_arch}
\vspace{3mm}
\end{figure*}

\subsection{Implementation Details}
 Our production architecture illustrated in Figure \ref{fig:prod_arch} features two components --- \textit{Near Real-Time (NRT)} Inference and \textit{Batch Inference}. Batch inference primarily serves items with a delay, whereas NRT serves items on an urgent basis, such as items newly created or revised by sellers. The batch inference is done in two parts: 1) for all items and keyphrases, and 2) daily differential (Diff), i.e. the difference of all new items and keyphrases created/revised and then merged with the existing ones. The NRT inference for our relevance model is written in the service code using DeepJava \cite{Deepjavalibrary} and onnx \cite{onnxruntime} serving. The NRT service is triggered by the event of creation or revision, behind a Flink processing window and feature enrichment modules.  

The process of full batch inference encompasses approximately 2.3 billion items, 1 billion keyphrases, and nearly 100 billion combinations of items and keyphrases. In contrast, daily differential processing involves servicing about 20 million items, 3 million keyphrases, and approximately 7 billion item-keyphrase pairings. Since the full batch inference is executed once, it is the latency of the differential (Diff) inference that determines the feasibility of deploying a model into production. The latency for the Diff inference is approximately 1.5 hours. All latency measurements were obtained using PySpark \cite{spark} with a configuration of 1000 executors, 20GB of memory, and 4 CPU cores, in conjunction with the transformers library \cite{wolf-etal-2020-transformers} and onnxruntime \cite{onnxruntime}. Although the possibility of employing GPU resources was explored, it proved to be financially prohibitive. 

\subsection{Impact}

In our recent deployment, we introduced the LLM (+/-) model into production, and ran an A/B test with the Search (+/-) model for 2 weeks across all major marketplaces. As a consequence, we observed a $15\%$ decrease in the number of keyphrases delivered to sellers (normalized by the item count).\footnote{This is the decrease at the adgroup level. Our serving of keyphrases is deduplicated by the formation of an adgroup, i.e. collection of items. Thus the deduplication effect is dampened at an adgroup level. Hence we see a dampened effect in comparison to the offline analysis which is at an item level. We cannot simulate adgroup analytics offline as it is formed at the time of presentation to the seller.} Furthermore, there was a $4\%$ enhancement in the adoption rate, defined as the proportion of keyphrases accepted by sellers relative to the total number of suggested keyphrases --- showing a greater acceptance of our keyphrases by the sellers. Our analysis did not reveal any statistically notable impact on key business indicators such as revenue, gross merchandise volume bought (GMB), clicks, impressions or sales. However, there was a positive influence on certain ratio metrics: the click-through rate ($CTR = \frac{clicks}{impressions}$) increased by $2.5\%$, the conversion rate ($CVR = \frac{purchases}{clicks}$) improved by $5.3\%$, and the return on advertising spend ($ROAS = \frac{GMB}{Ads\;Revenue}$),  experienced a notable improvement of $11. 76\%$. 

Although A/B tests are a great way of measuring short-term effects, a more concrete long-term effect was reflected in our Seller sentiment survey. The seller sentiment surveys are conducted biannually, where sellers are polled about their sentiments about our product offerings and how they feel about certain aspects of our product. Following the release of our new relevance model, we observed a noteworthy decline in the dissatisfaction rate concerning the relevance of our keyphrase suggestions from $6\%$ to $3\%$ in the subsequent quarter following the launch of the model. This progression underscores the substantial and enduring influence of our model in enhancing seller satisfaction.

\definecolor{green1}{RGB}{30,70,30}     
\definecolor{green2}{RGB}{80,130,80}
\definecolor{green3}{RGB}{140,190,140}
\definecolor{green4}{RGB}{200,230,200}
\definecolor{green5}{RGB}{235,250,235}  

\section{Conclusion}
This study conceptualizes Advertiser keyphrase relevance as the intricate interplay among three dynamical systems: the seller's judgment, impacting the adoption of our product; Advertising, which offers the keyphrases for bidding; and Search, which conducts auctions for these keyphrases. We present a distillation framework for training tiny cross-encoder networks from general LLMs like Mixtral 8x7B Instruct v0.1 and show it's effectiveness in emulating human judgment while still maintaining an alignment with Search judgment in the context of Advertiser keyphrase relevance. We also discuss the practical challenges of collecting reliable human judgment for a LLM to fine-tune on, thus paving the way for general LLMs as tools for data augmentation and evaluation. While we champion the use of general LLMs in this role, we also advocate for benchmarking said LLMs against some form of business metrics at scale and also present a framework for evaluation of models reliant on LLM generated relevance signals for advertiser keyphrase relevance grounded in business metrics. More generally, this study presents a discussion regarding various sources of ground-truth one can use to mitigate the various biases present in Advertising data and how we can leverage LLMs to mitigate these biases in a scalable manner in order to serve millions of users with an inventory of over 2.3 billion items.

\clearpage
\bibliography{mybibfile}

\begin{thebibliography}{44}
\providecommand{\natexlab}[1]{#1}
\providecommand{\url}[1]{\texttt{#1}}
\expandafter\ifx\csname urlstyle\endcsname\relax
  \providecommand{\doi}[1]{doi: #1}\else
  \providecommand{\doi}{doi: \begingroup \urlstyle{rm}\Url}\fi

\bibitem[Bavaresco et~al.(2024)Bavaresco, Bernardi, Bertolazzi, Elliott, Fernández, Gatt, Ghaleb, Giulianelli, Hanna, Koller, Martins, Mondorf, Neplenbroek, Pezzelle, Plank, Schlangen, Suglia, Surikuchi, Takmaz, and Testoni]{bavaresco2024llmsinsteadhumanjudges}
A.~Bavaresco, R.~Bernardi, L.~Bertolazzi, D.~Elliott, R.~Fernández, A.~Gatt, E.~Ghaleb, M.~Giulianelli, M.~Hanna, A.~Koller, A.~F.~T. Martins, P.~Mondorf, V.~Neplenbroek, S.~Pezzelle, B.~Plank, D.~Schlangen, A.~Suglia, A.~K. Surikuchi, E.~Takmaz, and A.~Testoni.
\newblock Llms instead of human judges? a large scale empirical study across 20 nlp evaluation tasks, 2024.
\newblock URL \url{https://arxiv.org/abs/2406.18403}.

\bibitem[Bhargava et~al.(2021)Bhargava, Drozd, and Rogers]{bhargava2021generalization}
P.~Bhargava, A.~Drozd, and A.~Rogers.
\newblock Generalization in nli: Ways (not) to go beyond simple heuristics, 2021.

\bibitem[Chen et~al.(2023)Chen, Dong, Wang, Feng, Wang, and He]{surveybias}
J.~Chen, H.~Dong, X.~Wang, F.~Feng, M.~Wang, and X.~He.
\newblock Bias and debias in recommender system: A survey and future directions.
\newblock \emph{ACM Trans. Inf. Syst.}, 41\penalty0 (3), Feb. 2023.
\newblock ISSN 1046-8188.
\newblock \doi{10.1145/3564284}.
\newblock URL \url{https://doi.org/10.1145/3564284}.

\bibitem[Deepjavalibrary()]{Deepjavalibrary}
Deepjavalibrary.
\newblock Deepjavalibrary/djl: An engine-agnostic deep learning framework in java.
\newblock URL \url{https://github.com/deepjavalibrary/djl}.

\bibitem[Deffayet et~al.(2023)Deffayet, Hager, Renders, and de~Rijke]{debiasedness}
R.~Deffayet, P.~Hager, J.-M. Renders, and M.~de~Rijke.
\newblock An offline metric for the debiasedness of click models.
\newblock In \emph{Proceedings of the 46th International ACM SIGIR Conference on Research and Development in Information Retrieval}, SIGIR '23, page 558–568, New York, NY, USA, 2023. Association for Computing Machinery.
\newblock ISBN 9781450394086.
\newblock \doi{10.1145/3539618.3591639}.
\newblock URL \url{https://doi.org/10.1145/3539618.3591639}.

\bibitem[developers(2021)]{onnxruntime}
O.~R. developers.
\newblock Onnx runtime.
\newblock \url{https://onnxruntime.ai/}, 2021.
\newblock Version: 1.20.1.

\bibitem[Dey et~al.(2024)Dey, An, and Levitan]{dey24_speechprosody}
S.~Dey, G.~An, and S.~I. Levitan.
\newblock Analysis and modeling of self-reported and observer-reported personality scores from text and speech.
\newblock In \emph{Speech Prosody 2024}, pages 975--979, 2024.
\newblock \doi{10.21437/SpeechProsody.2024-197}.

\bibitem[Dey et~al.(2025)Dey, Zhang, Wu, Dong, and Li]{dey2025middlemanbiasadvertisingaligning}
S.~Dey, W.~Zhang, H.~Wu, B.~Dong, and B.~Li.
\newblock Middleman bias in advertising: Aligning relevance of keyphrase recommendations with search.
\newblock In \emph{Companion Proceedings of the ACM on Web Conference 2025}, WWW '25, page 2701–2705, New York, NY, USA, 2025. Association for Computing Machinery.
\newblock ISBN 9798400713316.
\newblock \doi{10.1145/3701716.3717858}.
\newblock URL \url{https://doi.org/10.1145/3701716.3717858}.

\bibitem[eBay Advertising(2025)]{eBay}
eBay Advertising.
\newblock Promoted listings priority, Jan 2025.
\newblock URL \url{https://www.ebayadvertising.com/en/ad-solutions/ad-types/promoted-listings-priority/}.

\bibitem[Gao et~al.(2023)Gao, Han, Zhu, Yang, Jiang, Xu, and Zheng]{rec4ad}
J.~Gao, S.~Han, H.~Zhu, S.~Yang, Y.~Jiang, J.~Xu, and B.~Zheng.
\newblock Rec4ad: A free lunch to mitigate sample selection bias for ads ctr prediction in taobao.
\newblock In \emph{Proceedings of the 32nd ACM International Conference on Information and Knowledge Management}, CIKM '23, page 4574–4580, New York, NY, USA, 2023. Association for Computing Machinery.
\newblock ISBN 9798400701245.
\newblock \doi{10.1145/3583780.3615496}.
\newblock URL \url{https://doi.org/10.1145/3583780.3615496}.

\bibitem[Gu et~al.(2025)Gu, Jiang, Shi, Tan, Zhai, Xu, Li, Shen, Ma, Liu, Wang, Zhang, Wang, Gao, Ni, and Guo]{gu2025surveyllmasajudge}
J.~Gu, X.~Jiang, Z.~Shi, H.~Tan, X.~Zhai, C.~Xu, W.~Li, Y.~Shen, S.~Ma, H.~Liu, S.~Wang, K.~Zhang, Y.~Wang, W.~Gao, L.~Ni, and J.~Guo.
\newblock A survey on llm-as-a-judge, 2025.
\newblock URL \url{https://arxiv.org/abs/2411.15594}.

\bibitem[Guo et~al.(2023)Guo, Nie, Cheng, Cheng, Kankanhalli, and Del~Bimbo]{10.1145/3565266}
Y.~Guo, L.~Nie, H.~Cheng, Z.~Cheng, M.~Kankanhalli, and A.~Del~Bimbo.
\newblock On modality bias recognition and reduction.
\newblock \emph{ACM Trans. Multimedia Comput. Commun. Appl.}, 19\penalty0 (3), Feb. 2023.
\newblock ISSN 1551-6857.
\newblock \doi{10.1145/3565266}.
\newblock URL \url{https://doi.org/10.1145/3565266}.

\bibitem[Gurjar et~al.(2025)Gurjar, Liu, Kolli, Kumar, and Rahurkar]{gurjar2025dashclipleveragingmultimodalmodels}
O.~Gurjar, K.~S. Liu, P.~Kolli, U.~Kumar, and M.~Rahurkar.
\newblock Dashclip: Leveraging multimodal models for generating semantic embeddings for doordash, 2025.
\newblock URL \url{https://arxiv.org/abs/2504.07110}.

\bibitem[Huang et~al.(2024)Huang, Qu, Bu, Zhou, Liu, Yang, Xu, and Zhao]{huang2024empiricalstudyllmasajudgellm}
H.~Huang, Y.~Qu, X.~Bu, H.~Zhou, J.~Liu, M.~Yang, B.~Xu, and T.~Zhao.
\newblock An empirical study of llm-as-a-judge for llm evaluation: Fine-tuned judge model is not a general substitute for gpt-4, 2024.
\newblock URL \url{https://arxiv.org/abs/2403.02839}.

\bibitem[Jiang et~al.(2024)Jiang, Sablayrolles, Roux, Mensch, Savary, Bamford, Chaplot, de~las Casas, Hanna, Bressand, Lengyel, Bour, Lample, Lavaud, Saulnier, Lachaux, Stock, Subramanian, Yang, Antoniak, Scao, Gervet, Lavril, Wang, Lacroix, and Sayed]{jiang2024mixtralexperts}
A.~Q. Jiang, A.~Sablayrolles, A.~Roux, A.~Mensch, B.~Savary, C.~Bamford, D.~S. Chaplot, D.~de~las Casas, E.~B. Hanna, F.~Bressand, G.~Lengyel, G.~Bour, G.~Lample, L.~R. Lavaud, L.~Saulnier, M.-A. Lachaux, P.~Stock, S.~Subramanian, S.~Yang, S.~Antoniak, T.~L. Scao, T.~Gervet, T.~Lavril, T.~Wang, T.~Lacroix, and W.~E. Sayed.
\newblock Mixtral of experts, 2024.
\newblock URL \url{https://arxiv.org/abs/2401.04088}.

\bibitem[Joachims et~al.(2007)Joachims, Granka, Pan, Hembrooke, Radlinski, and Gay]{clicksNquery}
T.~Joachims, L.~Granka, B.~Pan, H.~Hembrooke, F.~Radlinski, and G.~Gay.
\newblock Evaluating the accuracy of implicit feedback from clicks and query reformulations in web search.
\newblock \emph{ACM Trans. Inf. Syst.}, 25\penalty0 (2):\penalty0 7–es, apr 2007.
\newblock ISSN 1046-8188.
\newblock \doi{10.1145/1229179.1229181}.
\newblock URL \url{https://doi.org/10.1145/1229179.1229181}.

\bibitem[Joachims et~al.(2017)Joachims, Swaminathan, and Schnabel]{learning2rank}
T.~Joachims, A.~Swaminathan, and T.~Schnabel.
\newblock Unbiased learning-to-rank with biased feedback.
\newblock In \emph{Proceedings of the Tenth ACM International Conference on Web Search and Data Mining}, WSDM '17, page 781–789, New York, NY, USA, 2017. Association for Computing Machinery.
\newblock ISBN 9781450346757.
\newblock \doi{10.1145/3018661.3018699}.
\newblock URL \url{https://doi.org/10.1145/3018661.3018699}.

\bibitem[Li et~al.(2023)Li, Zhang, Dubois, Taori, Gulrajani, Guestrin, Liang, and Hashimoto]{alpaca_eval}
X.~Li, T.~Zhang, Y.~Dubois, R.~Taori, I.~Gulrajani, C.~Guestrin, P.~Liang, and T.~B. Hashimoto.
\newblock Alpacaeval: An automatic evaluator of instruction-following models.
\newblock \url{https://github.com/tatsu-lab/alpaca_eval}, 5 2023.

\bibitem[Lichtenberg et~al.(2024)Lichtenberg, Buchholz, and Schwöbel]{lichtenberg2024largelanguagemodelsrecommender}
J.~M. Lichtenberg, A.~Buchholz, and P.~Schwöbel.
\newblock Large language models as recommender systems: A study of popularity bias, 2024.
\newblock URL \url{https://arxiv.org/abs/2406.01285}.

\bibitem[Lim et~al.(2015)Lim, McAuley, and Lanckriet]{lim2015top}
D.~Lim, J.~McAuley, and G.~Lanckriet.
\newblock Top-n recommendation with missing implicit feedback.
\newblock In \emph{Proceedings of the 9th ACM Conference on Recommender Systems}, pages 309--312, 2015.

\bibitem[Mishra et~al.(2024)Mishra, Dey, Zhao, Wu, Li, and Madduri]{ashirbad-etal-2024}
A.~Mishra, S.~Dey, J.~Zhao, H.~Wu, B.~Li, and K.~Madduri.
\newblock Graphite: A graph-based extreme multi-label short text classifier for keyphrase recommendation.
\newblock In \emph{{27th European Conference on Artificial Intelligence, 19–24 October 2024, Santiago de Compostela, Spain – Including 13th Conference on Prestigious Applications of Intelligent Systems (PAIS 2024)}}, volume 392 of \emph{Frontiers in Artificial Intelligence and Applications}, pages 4657 -- 4664. {IOS} Press, 2024.
\newblock URL \url{https://doi.org/10.3233/FAIA241061}.

\bibitem[Mishra et~al.(2025)Mishra, Dey, Wu, Zhao, Yu, Ni, Li, and Madduri]{mishra2024graphexgraphbasedextractionmethod}
A.~Mishra, S.~Dey, H.~Wu, J.~Zhao, H.~Yu, K.~Ni, B.~Li, and K.~Madduri.
\newblock { GraphEx: A Graph-Based Extraction Method for Advertiser Keyphrase Recommendation }.
\newblock In \emph{2025 IEEE 41st International Conference on Data Engineering (ICDE)}, pages 4400--4413, Los Alamitos, CA, USA, May 2025. IEEE Computer Society.
\newblock \doi{10.1109/ICDE65448.2025.00330}.
\newblock URL \url{https://doi.ieeecomputersociety.org/10.1109/ICDE65448.2025.00330}.

\bibitem[Murphy(1996)]{jaccard}
A.~H. Murphy.
\newblock The finley affair: A signal event in the history of forecast verification.
\newblock \emph{Weather and Forecasting}, 11\penalty0 (1):\penalty0 3 -- 20, 1996.
\newblock \doi{10.1175/1520-0434(1996)011<0003:TFAASE>2.0.CO;2}.
\newblock URL \url{https://journals.ametsoc.org/view/journals/wefo/11/1/1520-0434_1996_011_0003_tfaase_2_0_co_2.xml}.

\bibitem[OpenAI et~al.(2024)OpenAI, Achiam, Adler, Agarwal, Ahmad, Akkaya, Aleman, Almeida, Altenschmidt, Altman, Anadkat, Avila, Babuschkin, Balaji, Balcom, Baltescu, Bao, Bavarian, Belgum, Bello, Berdine, Bernadett-Shapiro, Berner, Bogdonoff, Boiko, Boyd, Brakman, Brockman, Brooks, Brundage, Button, Cai, Campbell, Cann, Carey, Carlson, Carmichael, Chan, Chang, Chantzis, Chen, Chen, Chen, Chen, Chen, Chess, Cho, Chu, Chung, Cummings, Currier, Dai, Decareaux, Degry, Deutsch, Deville, Dhar, Dohan, Dowling, Dunning, Ecoffet, Eleti, Eloundou, Farhi, Fedus, Felix, Fishman, Forte, Fulford, Gao, Georges, Gibson, Goel, Gogineni, Goh, Gontijo-Lopes, Gordon, Grafstein, Gray, Greene, Gross, Gu, Guo, Hallacy, Han, Harris, He, Heaton, Heidecke, Hesse, Hickey, Hickey, Hoeschele, Houghton, Hsu, Hu, Hu, Huizinga, Jain, Jain, Jang, Jiang, Jiang, Jin, Jin, Jomoto, Jonn, Jun, Kaftan, Łukasz Kaiser, Kamali, Kanitscheider, Keskar, Khan, Kilpatrick, Kim, Kim, Kim, Kirchner, Kiros, Knight, Kokotajlo, Łukasz Kondraciuk, Kondrich,
  Konstantinidis, Kosic, Krueger, Kuo, Lampe, Lan, Lee, Leike, Leung, Levy, Li, Lim, Lin, Lin, Litwin, Lopez, Lowe, Lue, Makanju, Malfacini, Manning, Markov, Markovski, Martin, Mayer, Mayne, McGrew, McKinney, McLeavey, McMillan, McNeil, Medina, Mehta, Menick, Metz, Mishchenko, Mishkin, Monaco, Morikawa, Mossing, Mu, Murati, Murk, Mély, Nair, Nakano, Nayak, Neelakantan, Ngo, Noh, Ouyang, O'Keefe, Pachocki, Paino, Palermo, Pantuliano, Parascandolo, Parish, Parparita, Passos, Pavlov, Peng, Perelman, de~Avila Belbute~Peres, Petrov, de~Oliveira~Pinto, Michael, Pokorny, Pokrass, Pong, Powell, Power, Power, Proehl, Puri, Radford, Rae, Ramesh, Raymond, Real, Rimbach, Ross, Rotsted, Roussez, Ryder, Saltarelli, Sanders, Santurkar, Sastry, Schmidt, Schnurr, Schulman, Selsam, Sheppard, Sherbakov, Shieh, Shoker, Shyam, Sidor, Sigler, Simens, Sitkin, Slama, Sohl, Sokolowsky, Song, Staudacher, Such, Summers, Sutskever, Tang, Tezak, Thompson, Tillet, Tootoonchian, Tseng, Tuggle, Turley, Tworek, Uribe, Vallone, Vijayvergiya,
  Voss, Wainwright, Wang, Wang, Wang, Ward, Wei, Weinmann, Welihinda, Welinder, Weng, Weng, Wiethoff, Willner, Winter, Wolrich, Wong, Workman, Wu, Wu, Wu, Xiao, Xu, Yoo, Yu, Yuan, Zaremba, Zellers, Zhang, Zhang, Zhao, Zheng, Zhuang, Zhuk, and Zoph]{openai2024gpt4technicalreport}
OpenAI, J.~Achiam, S.~Adler, S.~Agarwal, L.~Ahmad, I.~Akkaya, F.~L. Aleman, D.~Almeida, J.~Altenschmidt, S.~Altman, S.~Anadkat, R.~Avila, I.~Babuschkin, S.~Balaji, V.~Balcom, P.~Baltescu, H.~Bao, M.~Bavarian, J.~Belgum, I.~Bello, J.~Berdine, G.~Bernadett-Shapiro, C.~Berner, L.~Bogdonoff, O.~Boiko, M.~Boyd, A.-L. Brakman, G.~Brockman, T.~Brooks, M.~Brundage, K.~Button, T.~Cai, R.~Campbell, A.~Cann, B.~Carey, C.~Carlson, R.~Carmichael, B.~Chan, C.~Chang, F.~Chantzis, D.~Chen, S.~Chen, R.~Chen, J.~Chen, M.~Chen, B.~Chess, C.~Cho, C.~Chu, H.~W. Chung, D.~Cummings, J.~Currier, Y.~Dai, C.~Decareaux, T.~Degry, N.~Deutsch, D.~Deville, A.~Dhar, D.~Dohan, S.~Dowling, S.~Dunning, A.~Ecoffet, A.~Eleti, T.~Eloundou, D.~Farhi, L.~Fedus, N.~Felix, S.~P. Fishman, J.~Forte, I.~Fulford, L.~Gao, E.~Georges, C.~Gibson, V.~Goel, T.~Gogineni, G.~Goh, R.~Gontijo-Lopes, J.~Gordon, M.~Grafstein, S.~Gray, R.~Greene, J.~Gross, S.~S. Gu, Y.~Guo, C.~Hallacy, J.~Han, J.~Harris, Y.~He, M.~Heaton, J.~Heidecke, C.~Hesse, A.~Hickey,
  W.~Hickey, P.~Hoeschele, B.~Houghton, K.~Hsu, S.~Hu, X.~Hu, J.~Huizinga, S.~Jain, S.~Jain, J.~Jang, A.~Jiang, R.~Jiang, H.~Jin, D.~Jin, S.~Jomoto, B.~Jonn, H.~Jun, T.~Kaftan, Łukasz Kaiser, A.~Kamali, I.~Kanitscheider, N.~S. Keskar, T.~Khan, L.~Kilpatrick, J.~W. Kim, C.~Kim, Y.~Kim, J.~H. Kirchner, J.~Kiros, M.~Knight, D.~Kokotajlo, Łukasz Kondraciuk, A.~Kondrich, A.~Konstantinidis, K.~Kosic, G.~Krueger, V.~Kuo, M.~Lampe, I.~Lan, T.~Lee, J.~Leike, J.~Leung, D.~Levy, C.~M. Li, R.~Lim, M.~Lin, S.~Lin, M.~Litwin, T.~Lopez, R.~Lowe, P.~Lue, A.~Makanju, K.~Malfacini, S.~Manning, T.~Markov, Y.~Markovski, B.~Martin, K.~Mayer, A.~Mayne, B.~McGrew, S.~M. McKinney, C.~McLeavey, P.~McMillan, J.~McNeil, D.~Medina, A.~Mehta, J.~Menick, L.~Metz, A.~Mishchenko, P.~Mishkin, V.~Monaco, E.~Morikawa, D.~Mossing, T.~Mu, M.~Murati, O.~Murk, D.~Mély, A.~Nair, R.~Nakano, R.~Nayak, A.~Neelakantan, R.~Ngo, H.~Noh, L.~Ouyang, C.~O'Keefe, J.~Pachocki, A.~Paino, J.~Palermo, A.~Pantuliano, G.~Parascandolo, J.~Parish, E.~Parparita,
  A.~Passos, M.~Pavlov, A.~Peng, A.~Perelman, F.~de~Avila Belbute~Peres, M.~Petrov, H.~P. de~Oliveira~Pinto, Michael, Pokorny, M.~Pokrass, V.~H. Pong, T.~Powell, A.~Power, B.~Power, E.~Proehl, R.~Puri, A.~Radford, J.~Rae, A.~Ramesh, C.~Raymond, F.~Real, K.~Rimbach, C.~Ross, B.~Rotsted, H.~Roussez, N.~Ryder, M.~Saltarelli, T.~Sanders, S.~Santurkar, G.~Sastry, H.~Schmidt, D.~Schnurr, J.~Schulman, D.~Selsam, K.~Sheppard, T.~Sherbakov, J.~Shieh, S.~Shoker, P.~Shyam, S.~Sidor, E.~Sigler, M.~Simens, J.~Sitkin, K.~Slama, I.~Sohl, B.~Sokolowsky, Y.~Song, N.~Staudacher, F.~P. Such, N.~Summers, I.~Sutskever, J.~Tang, N.~Tezak, M.~B. Thompson, P.~Tillet, A.~Tootoonchian, E.~Tseng, P.~Tuggle, N.~Turley, J.~Tworek, J.~F.~C. Uribe, A.~Vallone, A.~Vijayvergiya, C.~Voss, C.~Wainwright, J.~J. Wang, A.~Wang, B.~Wang, J.~Ward, J.~Wei, C.~Weinmann, A.~Welihinda, P.~Welinder, J.~Weng, L.~Weng, M.~Wiethoff, D.~Willner, C.~Winter, S.~Wolrich, H.~Wong, L.~Workman, S.~Wu, J.~Wu, M.~Wu, K.~Xiao, T.~Xu, S.~Yoo, K.~Yu, Q.~Yuan,
  W.~Zaremba, R.~Zellers, C.~Zhang, M.~Zhang, S.~Zhao, T.~Zheng, J.~Zhuang, W.~Zhuk, and B.~Zoph.
\newblock Gpt-4 technical report, 2024.
\newblock URL \url{https://arxiv.org/abs/2303.08774}.

\bibitem[Ouyang et~al.(2022)Ouyang, Wu, Jiang, Almeida, Wainwright, Mishkin, Zhang, Agarwal, Slama, Ray, Schulman, Hilton, Kelton, Miller, Simens, Askell, Welinder, Christiano, Leike, and Lowe]{NEURIPS2022_b1efde53}
L.~Ouyang, J.~Wu, X.~Jiang, D.~Almeida, C.~Wainwright, P.~Mishkin, C.~Zhang, S.~Agarwal, K.~Slama, A.~Ray, J.~Schulman, J.~Hilton, F.~Kelton, L.~Miller, M.~Simens, A.~Askell, P.~Welinder, P.~F. Christiano, J.~Leike, and R.~Lowe.
\newblock Training language models to follow instructions with human feedback.
\newblock In S.~Koyejo, S.~Mohamed, A.~Agarwal, D.~Belgrave, K.~Cho, and A.~Oh, editors, \emph{Advances in Neural Information Processing Systems}, volume~35, pages 27730--27744. Curran Associates, Inc., 2022.
\newblock URL \url{https://proceedings.neurips.cc/paper_files/paper/2022/file/b1efde53be364a73914f58805a001731-Paper-Conference.pdf}.

\bibitem[Park et~al.(2024)Park, Jang, Alasaly, Mopidevi, Zolensky, Eaton, Lee, and Johnson]{park2024assessingmodalitybiasvideo}
J.~Park, K.~J. Jang, B.~Alasaly, S.~Mopidevi, A.~Zolensky, E.~Eaton, I.~Lee, and K.~Johnson.
\newblock Assessing modality bias in video question answering benchmarks with multimodal large language models, 2024.
\newblock URL \url{https://arxiv.org/abs/2408.12763}.

\bibitem[research team()]{databricksIntroducingDBRX}
M.~research team.
\newblock {I}ntroducing {D}{B}{R}{X}: {A} {N}ew {S}tate-of-the-{A}rt {O}pen {L}{L}{M} --- databricks.com.
\newblock \url{https://www.databricks.com/blog/introducing-dbrx-new-state-art-open-llm}.
\newblock [Accessed 16-04-2025].

\bibitem[Soboroff(2025)]{Soboroff_2025}
I.~Soboroff.
\newblock Don’t use llms to make relevance judgments.
\newblock \emph{Information Retrieval Research}, 1\penalty0 (1):\penalty0 29–46, Mar. 2025.
\newblock \doi{10.54195/irrj.19625}.
\newblock URL \url{https://irrj.org/article/view/19625}.

\bibitem[Steck(2010)]{MNAR}
H.~Steck.
\newblock Training and testing of recommender systems on data missing not at random.
\newblock In \emph{Proceedings of the 16th ACM SIGKDD International Conference on Knowledge Discovery and Data Mining}, KDD '10, page 713–722, New York, NY, USA, 2010. Association for Computing Machinery.
\newblock ISBN 9781450300551.
\newblock \doi{10.1145/1835804.1835895}.
\newblock URL \url{https://doi.org/10.1145/1835804.1835895}.

\bibitem[Stureborg et~al.(2024)Stureborg, Alikaniotis, and Suhara]{stureborg2024largelanguagemodelsinconsistent}
R.~Stureborg, D.~Alikaniotis, and Y.~Suhara.
\newblock Large language models are inconsistent and biased evaluators, 2024.
\newblock URL \url{https://arxiv.org/abs/2405.01724}.

\bibitem[Thomas et~al.(2024)Thomas, Spielman, Craswell, and Mitra]{microsoft_llm_as_a_judge}
P.~Thomas, S.~Spielman, N.~Craswell, and B.~Mitra.
\newblock Large language models can accurately predict searcher preferences.
\newblock SIGIR '24, page 1930–1940, New York, NY, USA, 2024. Association for Computing Machinery.
\newblock ISBN 9798400704314.
\newblock \doi{10.1145/3626772.3657707}.
\newblock URL \url{https://doi.org/10.1145/3626772.3657707}.

\bibitem[Touvron et~al.(2023)Touvron, Martin, Stone, Albert, Almahairi, Babaei, Bashlykov, Batra, Bhargava, Bhosale, Bikel, Blecher, Ferrer, Chen, Cucurull, Esiobu, Fernandes, Fu, Fu, Fuller, Gao, Goswami, Goyal, Hartshorn, Hosseini, Hou, Inan, Kardas, Kerkez, Khabsa, Kloumann, Korenev, Koura, Lachaux, Lavril, Lee, Liskovich, Lu, Mao, Martinet, Mihaylov, Mishra, Molybog, Nie, Poulton, Reizenstein, Rungta, Saladi, Schelten, Silva, Smith, Subramanian, Tan, Tang, Taylor, Williams, Kuan, Xu, Yan, Zarov, Zhang, Fan, Kambadur, Narang, Rodriguez, Stojnic, Edunov, and Scialom]{touvron2023llama2openfoundation}
H.~Touvron, L.~Martin, K.~Stone, P.~Albert, A.~Almahairi, Y.~Babaei, N.~Bashlykov, S.~Batra, P.~Bhargava, S.~Bhosale, D.~Bikel, L.~Blecher, C.~C. Ferrer, M.~Chen, G.~Cucurull, D.~Esiobu, J.~Fernandes, J.~Fu, W.~Fu, B.~Fuller, C.~Gao, V.~Goswami, N.~Goyal, A.~Hartshorn, S.~Hosseini, R.~Hou, H.~Inan, M.~Kardas, V.~Kerkez, M.~Khabsa, I.~Kloumann, A.~Korenev, P.~S. Koura, M.-A. Lachaux, T.~Lavril, J.~Lee, D.~Liskovich, Y.~Lu, Y.~Mao, X.~Martinet, T.~Mihaylov, P.~Mishra, I.~Molybog, Y.~Nie, A.~Poulton, J.~Reizenstein, R.~Rungta, K.~Saladi, A.~Schelten, R.~Silva, E.~M. Smith, R.~Subramanian, X.~E. Tan, B.~Tang, R.~Taylor, A.~Williams, J.~X. Kuan, P.~Xu, Z.~Yan, I.~Zarov, Y.~Zhang, A.~Fan, M.~Kambadur, S.~Narang, A.~Rodriguez, R.~Stojnic, S.~Edunov, and T.~Scialom.
\newblock Llama 2: Open foundation and fine-tuned chat models, 2023.
\newblock URL \url{https://arxiv.org/abs/2307.09288}.

\bibitem[Turc et~al.(2019)Turc, Chang, Lee, and Toutanova]{bert-tiny}
I.~Turc, M.~Chang, K.~Lee, and K.~Toutanova.
\newblock Well-read students learn better: The impact of student initialization on knowledge distillation.
\newblock \emph{CoRR}, abs/1908.08962, 2019.
\newblock URL \url{http://arxiv.org/abs/1908.08962}.

\bibitem[Vella(1998{\natexlab{a}})]{sample_selection}
F.~Vella.
\newblock Estimating models with sample selection bias: A survey.
\newblock \emph{The Journal of Human Resources}, 33\penalty0 (1):\penalty0 127--169, 1998{\natexlab{a}}.
\newblock ISSN 0022166X.
\newblock URL \url{http://www.jstor.org/stable/146317}.

\bibitem[Vella(1998{\natexlab{b}})]{sampleselectionbias}
F.~Vella.
\newblock Estimating models with sample selection bias: A survey.
\newblock \emph{The Journal of Human Resources}, 33\penalty0 (1):\penalty0 127--169, 1998{\natexlab{b}}.
\newblock ISSN 0022166X.
\newblock URL \url{http://www.jstor.org/stable/146317}.

\bibitem[Wang et~al.(2024)Wang, Sundararaman, Gungor, Xu, Kamath, Chalasani, Hazra, and Rao]{wang2024improvingpinterestsearchrelevance}
H.~Wang, M.~N. Sundararaman, O.~Gungor, Y.~Xu, K.~Kamath, R.~Chalasani, K.~S. Hazra, and J.~Rao.
\newblock Improving pinterest search relevance using large language models, 2024.
\newblock URL \url{https://arxiv.org/abs/2410.17152}.

\bibitem[Wolf et~al.(2020)Wolf, Debut, Sanh, Chaumond, Delangue, Moi, Cistac, Rault, Louf, Funtowicz, Davison, Shleifer, von Platen, Ma, Jernite, Plu, Xu, Scao, Gugger, Drame, Lhoest, and Rush]{wolf-etal-2020-transformers}
T.~Wolf, L.~Debut, V.~Sanh, J.~Chaumond, C.~Delangue, A.~Moi, P.~Cistac, T.~Rault, R.~Louf, M.~Funtowicz, J.~Davison, S.~Shleifer, P.~von Platen, C.~Ma, Y.~Jernite, J.~Plu, C.~Xu, T.~L. Scao, S.~Gugger, M.~Drame, Q.~Lhoest, and A.~M. Rush.
\newblock Transformers: State-of-the-art natural language processing.
\newblock In \emph{Proceedings of the 2020 Conference on Empirical Methods in Natural Language Processing: System Demonstrations}, pages 38--45, Online, Oct. 2020. Association for Computational Linguistics.
\newblock URL \url{https://www.aclweb.org/anthology/2020.emnlp-demos.6}.

\bibitem[Yang et~al.(2024)Yang, Yang, Hui, Zheng, Yu, Zhou, Li, Li, Liu, Huang, Dong, Wei, Lin, Tang, Wang, Yang, Tu, Zhang, Ma, Yang, Xu, Zhou, Bai, He, Lin, Dang, Lu, Chen, Yang, Li, Xue, Ni, Zhang, Wang, Peng, Men, Gao, Lin, Wang, Bai, Tan, Zhu, Li, Liu, Ge, Deng, Zhou, Ren, Zhang, Wei, Ren, Liu, Fan, Yao, Zhang, Wan, Chu, Liu, Cui, Zhang, Guo, and Fan]{yang2024qwen2technicalreport}
A.~Yang, B.~Yang, B.~Hui, B.~Zheng, B.~Yu, C.~Zhou, C.~Li, C.~Li, D.~Liu, F.~Huang, G.~Dong, H.~Wei, H.~Lin, J.~Tang, J.~Wang, J.~Yang, J.~Tu, J.~Zhang, J.~Ma, J.~Yang, J.~Xu, J.~Zhou, J.~Bai, J.~He, J.~Lin, K.~Dang, K.~Lu, K.~Chen, K.~Yang, M.~Li, M.~Xue, N.~Ni, P.~Zhang, P.~Wang, R.~Peng, R.~Men, R.~Gao, R.~Lin, S.~Wang, S.~Bai, S.~Tan, T.~Zhu, T.~Li, T.~Liu, W.~Ge, X.~Deng, X.~Zhou, X.~Ren, X.~Zhang, X.~Wei, X.~Ren, X.~Liu, Y.~Fan, Y.~Yao, Y.~Zhang, Y.~Wan, Y.~Chu, Y.~Liu, Z.~Cui, Z.~Zhang, Z.~Guo, and Z.~Fan.
\newblock Qwen2 technical report, 2024.
\newblock URL \url{https://arxiv.org/abs/2407.10671}.

\bibitem[Ye et~al.(2024)Ye, Wang, Huang, Chen, Zhang, Moniz, Gao, Geyer, Huang, Chen, Chawla, and Zhang]{ye2024justiceprejudicequantifyingbiases}
J.~Ye, Y.~Wang, Y.~Huang, D.~Chen, Q.~Zhang, N.~Moniz, T.~Gao, W.~Geyer, C.~Huang, P.-Y. Chen, N.~V. Chawla, and X.~Zhang.
\newblock Justice or prejudice? quantifying biases in llm-as-a-judge, 2024.
\newblock URL \url{https://arxiv.org/abs/2410.02736}.

\bibitem[Yue et~al.(2010)Yue, Patel, and Roehrig]{beyondPosBias}
Y.~Yue, R.~Patel, and H.~Roehrig.
\newblock Beyond position bias: examining result attractiveness as a source of presentation bias in clickthrough data.
\newblock In \emph{Proceedings of the 19th International Conference on World Wide Web}, WWW '10, page 1011–1018, New York, NY, USA, 2010. Association for Computing Machinery.
\newblock ISBN 9781605587998.
\newblock \doi{10.1145/1772690.1772793}.
\newblock URL \url{https://doi.org/10.1145/1772690.1772793}.

\bibitem[Zaharia et~al.(2016)Zaharia, Xin, Wendell, Das, Armbrust, Dave, Meng, Rosen, Venkataraman, Franklin, Ghodsi, Gonzalez, Shenker, and Stoica]{spark}
M.~Zaharia, R.~S. Xin, P.~Wendell, T.~Das, M.~Armbrust, A.~Dave, X.~Meng, J.~Rosen, S.~Venkataraman, M.~J. Franklin, A.~Ghodsi, J.~Gonzalez, S.~Shenker, and I.~Stoica.
\newblock Apache spark: a unified engine for big data processing.
\newblock \emph{Commun. ACM}, 59\penalty0 (11):\penalty0 56–65, Oct. 2016.
\newblock ISSN 0001-0782.
\newblock \doi{10.1145/2934664}.
\newblock URL \url{https://doi.org/10.1145/2934664}.

\bibitem[Zhang et~al.(2025)Zhang, U{\c{c}}ar, Dey, Wu, Li, and Zhang]{zhang2024lazyprolifictacklingmissing}
R.~H. Zhang, B.~U{\c{c}}ar, S.~Dey, H.~Wu, B.~Li, and R.~Zhang.
\newblock From lazy to prolific: Tackling missing labels in open vocabulary extreme classification by positive-unlabeled sequence learning.
\newblock In L.~Chiruzzo, A.~Ritter, and L.~Wang, editors, \emph{Findings of the Association for Computational Linguistics: NAACL 2025}, pages 1--16, Albuquerque, New Mexico, Apr. 2025. Association for Computational Linguistics.
\newblock ISBN 979-8-89176-195-7.
\newblock URL \url{https://aclanthology.org/2025.findings-naacl.1/}.

\bibitem[Zheng et~al.(2023)Zheng, Chiang, Sheng, Zhuang, Wu, Zhuang, Lin, Li, Li, Xing, Zhang, Gonzalez, and Stoica]{NEURIPS2023_91f18a12}
L.~Zheng, W.-L. Chiang, Y.~Sheng, S.~Zhuang, Z.~Wu, Y.~Zhuang, Z.~Lin, Z.~Li, D.~Li, E.~Xing, H.~Zhang, J.~E. Gonzalez, and I.~Stoica.
\newblock Judging llm-as-a-judge with mt-bench and chatbot arena.
\newblock In A.~Oh, T.~Naumann, A.~Globerson, K.~Saenko, M.~Hardt, and S.~Levine, editors, \emph{Advances in Neural Information Processing Systems}, volume~36, pages 46595--46623. Curran Associates, Inc., 2023.
\newblock URL \url{https://proceedings.neurips.cc/paper_files/paper/2023/file/91f18a1287b398d378ef22505bf41832-Paper-Datasets_and_Benchmarks.pdf}.

\bibitem[Zhu et~al.(2025)Zhu, Wang, and Wang]{zhu2025judgelmfinetunedlargelanguage}
L.~Zhu, X.~Wang, and X.~Wang.
\newblock Judgelm: Fine-tuned large language models are scalable judges, 2025.
\newblock URL \url{https://arxiv.org/abs/2310.17631}.

\end{thebibliography}

\end{document}